\documentclass[aps,prl,twocolumn,english,superscriptaddress,amsmath,amssymb,floatfix,longbibliography]{revtex4-2}
\usepackage[T1]{fontenc}
\usepackage{xcolor}
\usepackage{pdfcolmk}
\usepackage{babel}
\usepackage{amsmath}
\usepackage{float}
\usepackage{graphicx}
\usepackage{esint}

\PassOptionsToPackage{normalem}{ulem}
\usepackage{ulem}
\usepackage{braket}
\usepackage[unicode=true,pdfusetitle,
 bookmarks=true,bookmarksnumbered=false,bookmarksopen=false,
 breaklinks=false,pdfborder={0 0 0},pdfborderstyle={},backref=false,colorlinks=true]
 {hyperref}

\makeatletter

\providecolor{lyxadded}{rgb}{0,0,1}
\providecolor{lyxdeleted}{rgb}{1,0,0}

\DeclareRobustCommand{\lyxsout}[1]{\ifx\\#1\else\sout{#1}\fi}

\usepackage{dcolumn}
\usepackage{bm}
\usepackage{color}
\usepackage{soul}
\usepackage{babel}
\usepackage{amsfonts}
\usepackage{slashed}
\usepackage{enumerate}

\usepackage{babel}

\makeatother

\begin{document}
\global\long\def\sgn{\mathrm{sgn}}%
\global\long\def\ket#1{\left|#1\right\rangle }%
\global\long\def\bra#1{\left\langle #1\right|}%
\global\long\def\sp#1#2{\langle#1|#2\rangle}%
\global\long\def\abs#1{\left|#1\right|}%
\global\long\def\avg#1{\langle#1\rangle}%
\global\long\def\purity{P}%

\title{Interference of local-measurement histories}
\author{Parveen Kumar}
\affiliation{Department of Physics, Indian Institute of Technology Jammu, Jammu 181221, India}
\author{Igor V. Gornyi}
\affiliation{\mbox{Institute for Quantum Materials and Technologies, Karlsruhe Institute of Technology, 76131 Karlsruhe, Germany}}
\affiliation{\mbox{Institut für Theorie der Kondensierten Materie, Karlsruhe Institute of Technology, 76131 Karlsruhe, Germany}}
\author{Yuval Gefen}
\affiliation{Department of Condensed Matter Physics, Weizmann Institute of Science, Rehovot 7610001, Israel}

\begin{abstract}
The evolution of a quantum system comprises two fundamental processes—continuous unitary dynamics and stochastic measurement-induced jumps. The latter are often viewed as a source of decoherence. Can two histories of such an evolution, made up of local measurements, interfere with each other?  Here, we answer this question in the affirmative. A manifestation of this interference is the generation of entanglement between two parts of the system that are individually coupled to distinct detectors. Specifically, we develop a protocol in which two alternative local measurement processes act on a pair of qubits, and show how interference of histories is generated under coherent control, leading to entanglement. Furthermore, we find that averaging over the detectors’ readouts still results in an entangled (albeit not maximally entangled) state. Our results extend the notion of quantum interference beyond unitary evolution to genuinely measurement-driven dynamics, and identify limits on the generation of quantum correlations using interference of measurement histories.
\end{abstract}

\maketitle

\emph{Introduction.}---
Interference is a defining feature of quantum dynamics, yet it is widely regarded as incompatible with quantum measurements. The evolution of an open quantum system constitutes a \textit{history} that comprises two elementary ingredients: unitary dynamics and stochastic state updates by measurements~\cite{Jacobs01092006,Wiseman_Milburn_2009,Jacobs_2014,Jordan_Siddiqi_2024,PhysRevX.9.031009,Li2019,PhysRevB.101.104301,Fisher2022}. These are commonly viewed as fundamentally different in their relation to coherence and quantum correlations: unitary evolution supports interference, whereas measurements are associated with information gain and decoherence. This sharp distinction shapes the common wisdom regarding measurement-driven dynamics, in which local measurements suppress quantum correlations and preclude interference-based effects.

Measurement-induced evolution is often described in terms of possible quantum trajectories labeled by sequences of measurement outcomes. Since these outcomes correspond to orthogonal, classically distinguishable records, different records combine incoherently. It is therefore tempting to conclude that measurement histories cannot interfere. This conclusion, however, rests on an implicit assumption that a history is uniquely determined by its recorded readouts. When distinct physical processes lead to the same outcomes, their contributions to the system state remain coherent within that subset of histories. Interference between such processes sharing the same classical record is therefore fully compatible with the presence of measurement readouts—and constitutes a genuine quantum resource in its own right.

\begin{figure}
    \centering
    \includegraphics[width=1.0\columnwidth]{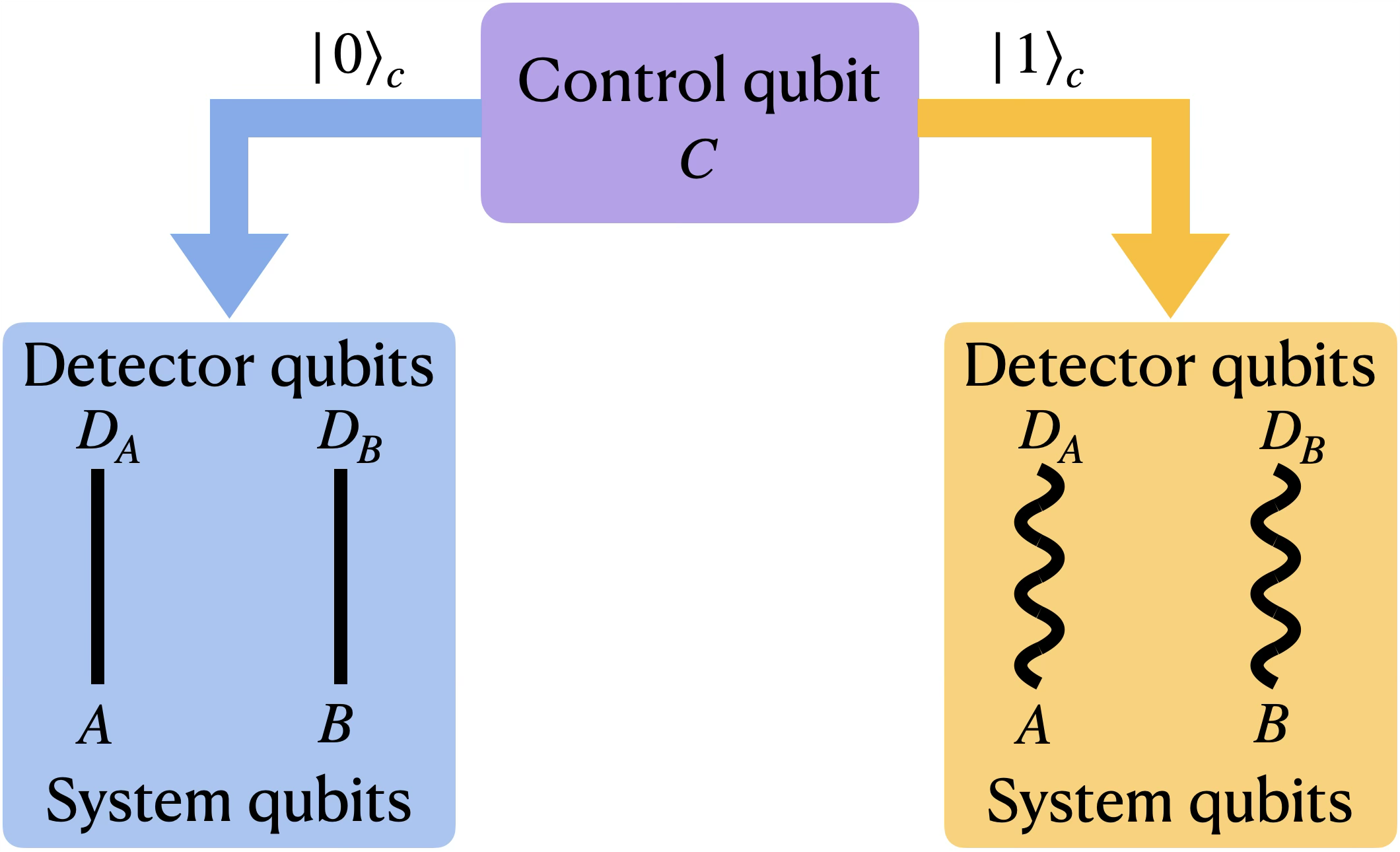} 
    \caption{Coherently controlled measurement histories. The control qubit $C$ selects between two alternative \textit{branches} of local measurements acting on system qubits $A$ and $B$. The left (right) branch corresponds to the measurement-induced history conditioned on $|0\rangle_C$ ($|1\rangle_C$). The respective solid and wiggly interaction lines denote that the local measurement Hamiltonians in the two branches may differ, corresponding to distinct measurement observables. After postselection of the control, amplitudes associated with the two histories interfere. This interference occurs within each fixed detector-readout sector, while different readout sectors remain orthogonal and combine incoherently.}
    \label{fig1:schematic}
\end{figure}

In this Letter, we put forward and establish the notion of \textit{interference of measurement-induced histories}. To demonstrate the physics, we analyze a minimal protocol in which two measurement processes—each consisting exclusively of local generalized measurements acting on separate subsystems—are placed in coherent superposition by an auxiliary control system, in the spirit of coherently controlled quantum operations~\cite{Chiribella2013,procopio2015experimental,PhysRevLett.117.100502,Rubino2017,PhysRevLett.120.120502,Felce2020,Chen2021b,Koudia2023,Goldberg2023,Rozema2024,Liu2025,zbgc-wdrk}. The essential structure of the coherently controlled measurement histories considered here is illustrated schematically in Fig.~\ref{fig1:schematic}. Individually, and in any classical mixture, these processes define local channels acting separately on the subsystems. Their interference, however, induces dynamics whose structure is resolved into sectors labeled by detector readouts. These readouts form orthogonal classical records that combine incoherently; yet within each fixed-readout sector the system state is a coherent superposition of the two measurement branches selected by the control. 

We show that interference of histories is not merely a formal possibility but can be harnessed to generate tangible effects. A direct and striking manifestation of this interference is the generation of entanglement---a phenomenon with no classical analogue~\cite{RevModPhys.81.865,AntonZeilinger_1998,erhard2020advances} that forms the foundation of quantum information processing~\cite{PhysRevLett.67.661,PhysRevLett.69.2881,PhysRevLett.70.1895,PhysRevLett.76.4656,bouwmeester1997experimental,PhysRevLett.80.1121,doi:10.1126/science.282.5389.706,PhysRevLett.80.3891,PhysRevLett.84.4729,PhysRevLett.86.5188,knill2001scheme}, and plays a central role in quantum many-body physics, including phase transitions~\cite{osterloh2002scaling,PhysRevA.66.032110,PhysRevLett.90.227902,PhysRevLett.92.027901,PhysRevA.73.042320,laflorencie2016quantum,PhysRevB.98.205136,PhysRevX.9.031009,PhysRevB.101.104301}---from an initially separable state. 
This is achieved using only local measurements, without joint operations involving both subsystems or any adaptive feedback~\cite{PhysRevB.67.241305,shankar2013autonomously,riste2013deterministic,PhysRevLett.112.170501,PhysRevA.92.062321,PhysRevA.102.062418,PhysRevA.107.032410,Herasymenko2023,Morales2024}. The interference-based entanglement generation persists even after unconditional averaging over all detector readouts—the natural default when readouts are discarded (as in blind, or passive, steering~\cite{PhysRevResearch.2.033347,PhysRevResearch.2.042014,PhysRevA.105.L010203,PhysRevResearch.6.023159}). We characterize entanglement quantitatively by computing concurrence analytically: before detector averaging, maximal entanglement can be reached; after averaging, entanglement survives but is fundamentally bounded.
These results establish interference of histories as a source of genuinely measurement-driven quantum dynamics, extending the notion of quantum coherence beyond unitary evolution.

\emph{Interference of Local Measurement Processes.}---
We consider a protocol in which two alternative local measurement processes act on a bipartite quantum system and are placed in coherent superposition, as illustrated in Fig.~\ref{fig1:schematic}. Two system qubits, labeled $A$ and $B$, are initialized in a separable state. An auxiliary control qubit $C$ governs the system-detector couplings: the $|0\rangle_C$ and $|1\rangle_C$ components of its state are associated with two distinct measurement histories, each implementing only local measurements on $A$ and $B$.
Following the system-detector interaction, the state of $C$ is postselected, while the measurement backaction on the system qubits may be retained, conditioned upon, or averaged over, depending on the operational setting considered.

Each evolution branch itself defines a local measurement channel. In particular, if the control does not exist, or alternatively one averages over (traces out) the control, and then averages over detector readouts within a single branch, the resulting map is a local completely positive trace-preserving map and therefore cannot generate entanglement from a separable input~\cite{nielsen2010quantum}. The essential distinction here is that the branch choice is coherent rather than classical. The system does not experience a stochastic mixture of local channels; instead, amplitudes associated with distinct measurement processes interfere prior to the final projection of the control qubit. As a result, the reduced system dynamics depends on relative phases and amplitudes that are absent in any classical mixture of local measurement maps.

The structure of the protocol becomes clear when one keeps track of the measurement readouts at an intermediate stage. Distinct measurement readouts of $A$ and $B$ correspond to orthogonal detector records. Consequently, after postselection on the control, the unnormalized system state can be written as a sum of orthogonal contributions,
\begin{equation}
\rho_{AB} = \sum_{i,j} |\psi_{ij}\rangle\langle\psi_{ij}|,
\label{eq:rho_generic}
\end{equation}
where each state vector $|\psi_{ij}\rangle$ is associated with a fixed pair of local readouts $(i,j)$. Interference occurs within each fixed-readout sector, while different readout sectors add incoherently because the corresponding detector records are orthogonal. This decomposition follows directly from standard measurement theory and does not rely on any particular microscopic model.

We now present the measurement dynamics employing Kraus operators~\cite{KRAUS1971311,nielsen2010quantum,Jacobs_2014,Wiseman_Milburn_2009,Jordan_Siddiqi_2024,Jacobs01092006}. The two branches (or histories) of system-detectors evolution are coherently selected by the control qubit: branch 0  corresponds to the evolution conditioned on the control qubit being in state $\ket{0}_c$, while branch 1 corresponds to the evolution conditioned on $\ket{1}_c$. The branch 0 is specified by the Kraus operators $K_i^A \otimes K_j^B$, and the branch 1 by 
$M_i^A \otimes M_j^B$, where $i$ and $j$ label the local measurement readouts of $D_A$ and $D_B$. If the initial state of the system is $|\phi_A\rangle \otimes |\phi_B\rangle$ and the control is prepared and postselected such that the two
histories contribute with amplitudes $\alpha$ and $\beta$, the state appearing in Eq.~\eqref{eq:rho_generic} takes the form
\begin{equation}
|\psi_{ij}\rangle =
\alpha (K_i^A \otimes K_j^B)|\phi_A\phi_B\rangle
+
\beta (M_i^A \otimes M_j^B)|\phi_A\phi_B\rangle .
\label{eq:psi_ij}
\end{equation}
A microscopic detector model that generates this controlled Kraus structure is presented in the Appendix~\ref{appendix-microscopic-model}.

Equation~\eqref{eq:psi_ij} highlights the dynamics leading to the interference of histories. Given the readouts $(i,j)$, the state $|\psi_{ij}\rangle$ is a coherent superposition of two product states, each associated with one measurement branch. For $\alpha,\beta\neq 0$, the state $|\psi_{ij}\rangle$ is entangled if and only if the two vectors in each of the pairs $\{K_i^A|\phi_A\rangle,\,M_i^A|\phi_A\rangle\}$ and $\{K_j^B|\phi_B\rangle,\,M_j^B|\phi_B\rangle\}$ are linearly independent, both in subsystem $A$ and $B$ (i.e., the state has Schmidt rank two). If either pair is colinear, $|\psi_{ij}\rangle$ factorizes. Thus, interference of histories acts as an entangling mechanism precisely when the two branches generate locally noncolinear contributions on both subsystems.

\emph{Entanglement from interference of histories.}---
To make this explicit, we consider a two-qubit example. The system is initialized in the separable state
\begin{equation}
|\psi^{\rm in}_{AB}\rangle = |+\rangle_A |+\rangle_B, 
\qquad 
|+\rangle = \frac{|0\rangle + |1\rangle}{\sqrt{2}}.
\end{equation}
The control qubit is prepared in $|+\rangle_C$ and, at the end, postselected in $|-\rangle_C$, so that the two histories interfere with equal amplitudes and opposite phase.

Each branch consists of local measurements. The Kraus operators corresponding to the measured observable $\sigma_{\hat{n}}=\hat{n}\cdot\vec{\sigma}$, where $\hat{n}$ is a unit vector specifying the measurement axis and $\vec{\sigma}=(\sigma_x,\sigma_y,\sigma_z)$ is the vector of Pauli matrices,
\begin{equation}
L_{\pm}(g,\hat n) = \sqrt{\frac{1+g}{2}}\,\Pi_{\pm}(\hat n) + \sqrt{\frac{1-g}{2}}\,\Pi_{\mp}(\hat n),
\label{eq:Kraus-L}
\end{equation}
where $g\in[0,1]$ controls the measurement strength and $\Pi_{\pm}(\hat n)$ are the projectors onto the eigenstates of $\sigma_{\hat n}$ with eigenvalues $\pm 1$. 

The Born-rule probabilities for obtaining outcome $i=\pm 1$ on a given subsystem $X\in\{A,B\}$ are
\begin{equation}
p_i^{X}=\langle \phi_X | \left(L_i^{X}\right)^\dagger L_i^{X} | \phi_X \rangle,
\end{equation}
where $|\phi_X\rangle$ denotes the input state of subsystem $X$ (in this example, $|\phi_X\rangle=|+\rangle$). For the bipartite system, the joint ($A$ and $B$) readout probability is
\begin{equation}
p_{ij}=\langle \phi_A\phi_B | (L_i^{A}\otimes L_j^{B})^\dagger (L_i^{A}\otimes L_j^{B}) | \phi_A\phi_B \rangle,
\end{equation}
which factorizes as $p_{ij}=p_i^{A}p_j^{B}$ within each branch, reflecting the locality of the measurements.

We choose the two branches such that branch $0$ performs the measurement of the observable $\sigma_x$ on $A$ and $\sigma_{\hat n_0}$ on $B$, while branch $1$ performs measurements of the observables $\sigma_{\hat n_1}$ on $A$ and $\sigma_x$ on $B$. The measurement axes $\hat n_0$ and $\hat n_1$ are given by
\begin{equation}
\hat n_0 = \cos\theta\,\hat x - \sin\theta\,\hat z,
\qquad
\hat n_1 = \cos\theta\,\hat x + \sin\theta\,\hat z,
\label{n0-and-n1}
\end{equation}
where $\theta\in[0,2\pi]$. Therefore, the Kraus operators for branch 0 become
\begin{equation}
K_{\pm}^A = L_{\pm}(g,\hat x),\quad K_{\pm}^B = L_{\pm}(g,\hat n_0)
\end{equation}
and for branch 1
\begin{equation}
M_{\pm}^A = L_{\pm}(g,\hat n_1),\quad M_{\pm}^B = L_{\pm}(g,\hat x)
\end{equation}
For each readout pair $(i,j)\in\{+,-\}^2$, postselection on the control yields the (unnormalized) conditional system state
\begin{equation}
|\psi_{ij}\rangle
=
\frac12
\Big[
(K_i^A\otimes K_j^B)
-
(M_i^A\otimes M_j^B)
\Big]
|++\rangle .
\label{psi-tilde-ij}
\end{equation}
For $(i,j)=(+,+)$ and $(-,-)$, the state $|\psi_{ij}\rangle$ is proportional to the Bell state, i.e.,
\begin{equation}
|\psi_{ij}\rangle\propto |\Phi^{-}\rangle = \frac{|00\rangle - |11\rangle}{\sqrt{2}},
\end{equation}
yielding unit concurrence, 
\begin{equation}
\mathcal{C}_{++} = \mathcal{C}_{--} = 1.
\end{equation}
Therefore, this example shows that interference of measurement histories may generate a maximally entangled system state, even though each branch consists only of local measurements and is individually non-entangling.

The joint probability of obtaining detector readout pair $(i,j)$ and successfully postselecting the control in $\ket{-}_C$ is $P_{ij}=\langle\psi_{ij}|\psi_{ij}\rangle$.
For the two sectors ($i=j$) that yield maximally entangled states, one obtains
\begin{eqnarray}
P_{jj} &=& (1+ j g)\frac{(1-\sqrt{1-g^2})\sin^2\theta}{16}.
\end{eqnarray}
Thus, the total probability of obtaining a maximally entangled Bell state is
\begin{equation}
P_{\rm Bell}=P_{++}+P_{--}
=
\frac{(1-\sqrt{1-g^2})\sin^2\theta}{8}.
\end{equation}
The remaining readout pairs $(+,-)$ and $(-,+)$ yield states $|\psi_{+-}\rangle$ and $|\psi_{-+}\rangle$ that are generically entangled but do not reduce to a Bell state; their contributions are included in the readout-averaged state analyzed below.

\emph{Entanglement under detector readout averaging.}---The entanglement discussed above arises at the level of fixed detector readouts, where interference of histories occurs within each readout sector. We now consider the case in which the detector readouts are not resolved, and the system state is obtained by averaging over all measurement readouts after postselection on the control qubit. The resulting system state has the form given in Eq.~(\ref{eq:rho_generic}), where each state vector $|\psi_{ij}\rangle$ corresponds to a fixed pair of local detector readouts. A natural question is whether maximal entanglement can still be generated. 

A maximally entangled two-qubit state must be pure. Hence, $\rho_{AB}$ can be maximally entangled only if all vectors $|\psi_{ij}\rangle$ are mutually proportional, imposing strong constraints on the measurement operators.
For the generalized measurements considered, we show in Appendix~\ref{app:avg_case_noncolinearity} that proportionality of all $|\psi_{ij}\rangle$ across readout sectors---required for purity of $\rho_{AB}$---enforces local colinearity of the branch-induced vectors on at least one subsystem. For $g\neq 0$, such colinearity occurs only when the input state of that subsystem is an eigenstate of the measured observable, in which case both branches act identically up to a scalar, the state factorizes across that subsystem, and entanglement is precluded. 

This leads to the following incompatibility. In the present setting, entanglement generated via interference requires local noncolinearity of the branch-induced states on both subsystems. In contrast, maximal entanglement after readout averaging requires global proportionality of the states $|\psi_{ij}\rangle$ across all readout sectors, which enforces local colinearity. Since these conditions are incompatible, maximal entanglement cannot be obtained after averaging over detector readouts.

We now return to the two-qubit example introduced in the last section. After postselection on the control and averaging over all measurement readouts, the normalized system state takes the form
\begin{equation}
\label{rho_AB_avg_regime}
\rho_{AB}
=
\frac{\sum_{i,j}
|\psi_{ij}\rangle
\langle\psi_{ij}|}{\textrm{Tr}[\sum_{i,j}
|\psi_{ij}\rangle
\langle\psi_{ij}|]},
\end{equation}
where the state $|\psi_{ij}\rangle$ is given in Eq.~(\ref{psi-tilde-ij}). Using Wootters' formalism~\cite{PhysRevLett.80.2245}, the concurrence of this state $\rho_{AB}$ is given by (see Appendix~\ref{app:concurrence})
\begin{equation}
\label{concurrence-avg-regime}
\mathcal{C}_{\rm avg}(g,\theta)
=
\frac{g^2(1+\cos\theta)}
{4(1+\sqrt{1-g^2})-g^2(1-\cos\theta)}.
\end{equation}
The concurrence attains its maximum value $\mathcal{C}_{\rm avg}=1/2$ in the limit $g=1$ and $\theta\to 0$. At exactly $\theta=0$, however, the two branches become identical, as the measurement axes $\hat n_0$ and $\hat n_1$ both reduce to $\hat x$. As a result, the postselected amplitude corresponding to projection of the control qubit onto $|-\rangle_C$ vanishes: the success probability of projecting the control to $\ket{-}_C$ is zero. The value $\mathcal{C}_{\rm avg}=1/2$ should therefore be understood as a limiting value obtained as $\theta\to 0$, rather than at $\theta=0$ itself. 

In the case of no measurements, i.e., $g=0$, the concurrence vanishes, as expected. The concurrence also vanishes for $\theta=\pi$. In this case, the measurement axes in the two branches coincide [cf. Eq.~(\ref{n0-and-n1})], so that both branches implement identical local measurements and no interference between the corresponding histories is generated, as in the $\theta=0$ limit. For any $g\neq 0$ and $\theta\neq \pi$, the averaged state $\rho_{AB}$ is entangled. In contrast to the postselected case, however, the averaged state is mixed, and its concurrence is strictly bounded below unity for all $g$ and $\theta$, consistent with the general arguments presented above.

\begin{figure}
    \centering
    \includegraphics[width=1.0\columnwidth]{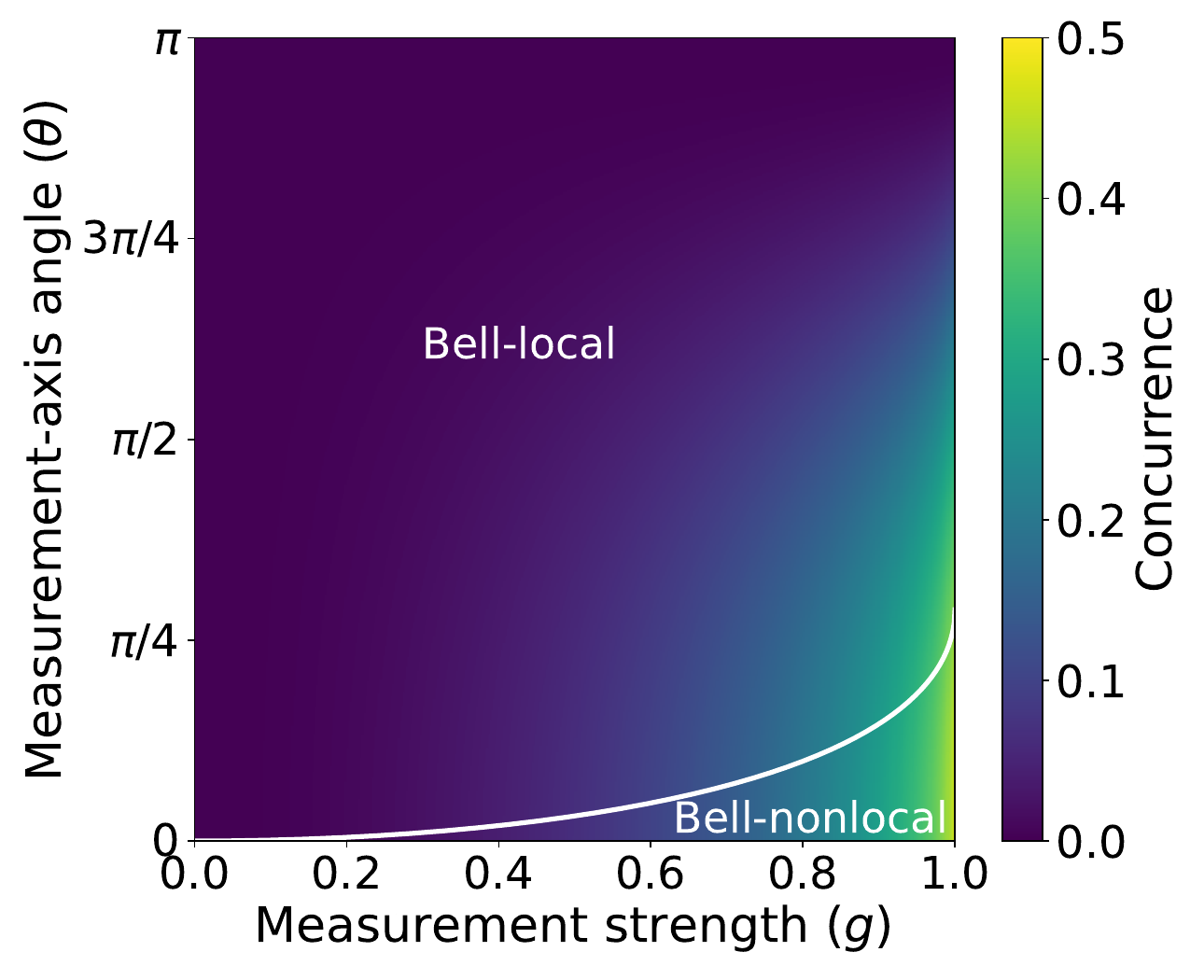} 
    \caption{Concurrence $\mathcal{C}_{\rm avg}(g,\theta)$ [cf. Eq.~(\ref{concurrence-avg-regime})] of the readout-averaged state for the two-qubit example, as a function of measurement strength $g$ and angle $\theta$. The solid white curve denotes the CHSH Bell-nonlocality boundary $\gamma(g,\theta)=1$, obtained from the Horodecki criterion for two-qubit states~\cite{HORODECKI1995340}. The figure shows two distinct regimes: an entangled but Bell-local region ($\gamma \leq 1$) and a Bell-nonlocal region ($\gamma > 1$), cf. Appendix~\ref{app:bell-boundary}. Although entanglement is generated over a broad parameter range, the concurrence remains strictly below unity for all $g$ and $\theta$, consistent with the constraint imposed by readout averaging.}
    \label{fig2:concurrence-heatmap}
\end{figure}

Let us now address the Bell nonlocality of the postselected and readout-averaged states. Here, by Bell nonlocality we refer to violation of the CHSH Bell inequality. In the postselected regime with measurement readout sequence $++$ or $--$, the system state is the pure Bell state $\ket{\Phi^-}$  with concurrence equal to one, and therefore following Gisin's theorem~\cite{GISIN1991201}, violates the Bell inequality. By contrast, after averaging over detector readouts, the state is mixed; entanglement then does not by itself guarantee Bell nonlocality, and the CHSH violation must be tested independently using the Horodecki criterion~\cite{HORODECKI1995340}. Using Eq.~(\ref{rho_AB_avg_regime}), one can write the correlation matrix $V_{mn}=\mathrm{Tr}[\rho_{AB}(\sigma_m\otimes\sigma_n)]$ and define $U=V^{\!T}V$. Let $\mu_1\ge\mu_2\ge\mu_3$ denote the eigenvalues of $U$. The maximal CHSH parameter is $B_{\max}=2\sqrt{\mu_1+\mu_2}$, and violation occurs iff $\gamma(g,\theta)\equiv \mu_1+\mu_2 > 1$ \cite{HORODECKI1995340}. Therefore, the Bell boundary for the state in our example is defined by the curve $\gamma(g,\theta)=1$ in the $(g,\theta)$ plane. The analytical expression for the Bell boundary is derived in Appendix~\ref{app:bell-boundary}. Figure~\ref{fig2:concurrence-heatmap} shows a heatmap of $\mathcal{C}_{\rm avg}(g,\theta)$ together with this boundary.  This reveals two distinct regimes in the averaged protocol:  entangled Bell-local ($\gamma\le1$) and Bell-nonlocal ($\gamma>1$) regions. This explicit example demonstrates that interference between local measurements can generate entanglement and Bell nonlocality, while maximal entanglement remains excluded under full readout averaging.

\emph{Discussion.}---We have shown that local measurement-induced histories can interfere when they are coherently controlled and share the same detector readout sequence. This establishes interference of histories as a feature of measurement-driven quantum dynamics, pushing it beyond the confines of unitary evolution. A clear manifestation of this interference is the generation of entanglement from an initially separable state, even though each measurement branch acts locally and is individually incapable of producing entanglement. Subjecting our protocol to averaging schemes, we identify the following constraint: for local generalized measurements, averaging over detector readouts precludes the generation of maximally entangled states. The readout-averaged states can nevertheless exhibit Bell nonlocality, as demonstrated explicitly in our two-qubit example. These results reveal both the robustness and the limitations of interference in local measurement-driven dynamics.

The origin of this behavior lies in the distinct roles of interference and averaging in measurement-driven dynamics. Interference between the two measurement branches occurs, at the level of fixed detector records, where the branch actions combine coherently and can act locally noncolinearly on both subsystems. One may then, optionally, average over different interference patterns. By contrast, averaging over readouts combines states associated with different detector records. Producing a maximally entangled state after such averaging would require the resulting state to be pure. Since readout averaging yields a weighted sum of pure state density matrices, obtaining this maximal entanglement is possible only if all readout-conditioned states are identical. For generalized local measurements, this requirement enforces local colinearity of the branch actions, eliminating the interference responsible for entanglement generation. The resulting constraint follows directly from the structure of the measurement protocol, without relying on fine-tuning or model-specific assumptions.

While conceptually related to the literature on coherently controlled quantum operations and indefinite causal order (ICO)~\cite{Chiribella2013,procopio2015experimental,PhysRevLett.117.100502,Rubino2017,PhysRevLett.120.120502,Felce2020,Chen2021b,Koudia2023,Goldberg2023,Rozema2024,Liu2025}, the present setting differs from that framework in two fundamental respects. First, the histories here are built exclusively from measurements—genuinely non-unitary, stochastic processes—rather than from unitary gates. This distinction is not merely technical: it directly challenges the conventional association of measurement readouts with decoherence and the suppression of quantum correlations. Second, prior ICO work has focused predominantly on superpositions in which the same pair of operations appears in reversed temporal order. The present framework is more general: the two histories may correspond to arbitrary, distinct local measurement processes, with ICO of measurements constituting only a narrow special case. We note that for unitary operations, the possibility of interfering histories beyond the ICO subclass was addressed previously~\cite{Chen2021b}. The present work differs in two essential ways: building blocks of the histories are measurements rather than unitaries, making the tension with decoherence the central conceptual issue; and entanglement generation is demonstrated even under full averaging over detector readouts, a regime entirely absent from the unitary framework of Ref.~\cite{Chen2021b}. The interplay between the classical detector record, the coherent measurement-branch superposition, and the readout-averaging procedure introduces a fascinating theme of measurement-induced interference dynamics.

Zooming in, from the conceptual formulation of interfering non-unitary histories to the facet of entanglement generation, our analysis highlights the interplay of coherent control and readout processing  (involving solely local measurements) in measurement-induced entanglement protocols. The latter complement earlier approaches based on joint measurements, explicit readout conditioning, or adaptive feedback~\cite{PhysRevB.67.241305,shankar2013autonomously,riste2013deterministic,PhysRevLett.112.170501,PhysRevA.92.062321,PhysRevA.102.062418,PhysRevA.107.032410,Herasymenko2023,Morales2024}, and specify when coherent control of local measurements provides an advantage and when restrictions arise under readout averaging. From an experimental perspective, the required ingredients are standard local measurements supplemented by coherent control over alternative operations, capabilities already demonstrated in quantum-switch experiments and interferometric control of quantum processes~\cite{procopio2015experimental,Rubino2017,PhysRevLett.121.090503,PhysRevLett.122.120504,PhysRevLett.131.240401,Rozema2024}. In particular, present protocol does not rely on nonlocal detectors or collective measurements, but only on extending such control techniques to local measurement settings. Our framework is naturally generalizable to include the combined effect of unitary evolution and measurements.

\emph{Conclusions and outlook.}---Our findings revise the conventional boundary between coherent and incoherent dynamics. The central lesson of this work is that the nature of a quantum process is not fixed by its constituent operations alone, but depends on how those operations are assembled into a history. Two measurement branches that are individually non-entangling, and remain so under any classical mixture, become entangling when combined coherently. It is only after averaging over the control (rather than over measurement readouts)  that phase relations are erased and a classical mixture of histories emerges, a phenomenon known as decoherence. Coherent control of measurement histories thus entails not merely a way of selecting operations, but a resource that changes what those operations can achieve -- one that is distinct from, and complementary to, both unitary coherence and joint-measurement protocols. 
Interference between measurement-induced histories thus emerges as an independent mechanism for generating and constraining quantum correlations.

This perspective elevates interference beyond unitary evolution into the domain of genuinely measurement-driven dynamics, with implications for quantum information, quantum optics, and the understanding of entanglement in many-body systems, including entanglement transitions. Several directions remain open: extensions to multipartite systems; to measurement classes beyond the generalized form considered here; to protocols combining coherent control with selective readout processing or adaptive feedback; and characterization of nonlocality across the full parameter space, including hidden Bell nonlocality~\cite{PhysRevLett.74.2619}. As a broader possible application, one can envisage a generalization of our protocol to Anderson localization phenomena driven by interference of long measurement histories.  
Taken together, these results provide a framework for understanding when interference of measurement histories can, and cannot, be used to generate quantum correlations---and what operational conditions determine this boundary.

\begin{acknowledgments}
We thank I. Poboiko for discussions. PK acknowledges support from the IIT Jammu Seed Grant (SGT-100106). The work was supported by the Deutsche Forschungsgemeinschaft (DFG,
German Research Foundation) through grants SH 81/8-1  and GO 1405/7-1. Y.G. also acknowledges support by the National Science Foundation (NSF)--Binational Science Foundation (BSF) through grant 2023666.

\end{acknowledgments}

\bibliography{references}

\appendix 





\section{Microscopic Model for Coherently Controlled Local Measurements}
\label{appendix-microscopic-model}
\subsection{General framework}
\label{gen-microscopic-model}

Here we present a microscopic model that generates the coherently controlled local measurement dynamics described in the main text. The total Hilbert space is
\begin{equation}
\mathcal{H}
=
\mathcal{H}_C\otimes\mathcal{H}_A\otimes\mathcal{H}_B
\otimes\mathcal{H}_{D_A}\otimes\mathcal{H}_{D_B},
\end{equation}
where $\mathcal{H}_C$ corresponds to the control qubit, $\mathcal{H}_{A,B}$ to the system qubits, and $\mathcal{H}_{D_A,D_B}$ to detector qubits locally coupled to $A$ and $B$. The initial joint state is taken as
\begin{equation}
|\Psi_{\mathrm{in}}\rangle
=
(\alpha|0\rangle_C+\beta|1\rangle_C)
\otimes|\phi_A\phi_B\rangle
\otimes|0\rangle_{D_A}|0\rangle_{D_B},
\end{equation}
with $|\alpha|^2+|\beta|^2=1$.

Local measurements are performed by coupling each system qubit $X\in\{A,B\}$ to its detector via a conditional interaction Hamiltonian
\begin{equation}
H_{SD}^X(\hat n)
=
\sum_{i=\pm}\Pi^{X}_i(\hat n)\otimes h^{X}_i ,
\end{equation}
where $\Pi^{X}_i(\hat n)$ are the projectors onto the eigenstates of $\hat n\cdot\vec{\sigma}$ acting on the system qubit, and $h^{X}_i$ are Hermitian operators acting on the detector. Evolution for a fixed interaction time $\tau$ gives
\begin{equation}
U_{SD}^X(\hat n)
=
e^{-i H_{SD}^X(\hat n)\tau}
=
\sum_{i=\pm}\Pi^{X}_i(\hat n)\otimes U^{X}_i ,
\label{eq:USD_general}
\end{equation}
with $U^{X}_i=e^{-i h^{X}_i\tau}$. Measurement of the detector in the computational basis $\{\ket{0},\ket{1}\}$ induces a backaction on the system qubit, and this backaction can be described using Kraus operators
\begin{eqnarray}
K_+^X &=&
{}_{D_X}\!\bra{0}\,U_{SD}^X(\hat n)\,\ket{0}_{D_X}, \\
K_-^X &=&
{}_{D_X}\!\bra{1}\,U_{SD}^X(\hat n)\,\ket{0}_{D_X}.
\end{eqnarray}
The Kraus operators satisfy the completeness condition, $\sum_i (K_i^X)^\dagger K_i^X=\mathbb I$. For suitable choices of $h_i^X$, the Kraus operators share a common eigenbasis and are characterized by a dimensionless measurement strength $g\in[0,1]$, interpolating between the identity map and a projective measurement limit.

The coherently controlled protocol can be generated by a single microscopic Hamiltonian acting on the joint control–system–detector Hilbert space. Introducing branch-dependent system–detector Hamiltonians $H_{SD}^{X,(0)}$ and $H_{SD}^{X,(1)}$, the total interaction Hamiltonian is
\begin{align}
H_{\mathrm{tot}}
&=
|0\rangle\!\langle0|_C\otimes
\bigl(H_{SD}^{A,(0)}\otimes\mathbb I_{D_B}
+
\mathbb I_{D_A}\otimes H_{SD}^{B,(0)}\bigr)
\nonumber\\
&+
|1\rangle\!\langle1|_C\otimes
\bigl(H_{SD}^{A,(1)}\otimes\mathbb I_{D_B}
+
\mathbb I_{D_A}\otimes H_{SD}^{B,(1)}\bigr).
\label{eq:Htot_SM}
\end{align}
Because the two control sectors are orthogonal, different terms in $H_{\mathrm{tot}}$ commute, and the corresponding unitary evolution is represented exactly as
\begin{align}
&U_{\mathrm{tot}} = e^{-iH_{\mathrm{tot}}\tau} \nonumber \\
&= |0\rangle\!\langle0|_C\otimes U_{SD}^{A,(0)}\otimes U_{SD}^{B,(0)} +
|1\rangle\!\langle1|_C\otimes U_{SD}^{A,(1)}\otimes U_{SD}^{B,(1)},
\label{eq:controlled_U_SM}
\end{align}
which is the controlled evolution used in the main text.

Measuring the detector qubits produces classical readouts $(i,j)$. Postselection on the control produces the unnormalized conditional system state
\begin{equation}
|\psi_{ij}\rangle
=
\alpha\,(K_i^A\otimes K_j^B)|\phi_A\phi_B\rangle
+
\beta\,(M_i^A\otimes M_j^B)|\phi_A\phi_B\rangle,
\end{equation}
while summing over detector readouts gives the unnormalized averaged state, 
$\rho_{AB}=\sum_{i,j}|\psi_{ij}\rangle\langle\psi_{ij}|$, which is Eq.~\eqref{eq:rho_generic} of the main text.

\subsection{Microscopic model for the canonical example of the main text}
\label{microscopic-model-for-example}

We now specialize the general microscopic construction of Sec.~\ref{gen-microscopic-model} to the two–qubit example discussed in the main text.  Our goal is to realize the Kraus operators~\eqref{eq:Kraus-L}
\begin{equation}
L_{\pm}(g,\hat n)
=
a\,\Pi_{\pm}(\hat n)
+
b\,\Pi_{\mp}(\hat n),
\label{eq:SM_Luders_form}
\end{equation}
with
\begin{equation}
a=\sqrt{\frac{1+g}{2}},\quad\textrm{and}\quad
b=\sqrt{\frac{1-g}{2}}.
\label{eq:a-and-b}
\end{equation}
Here, $\Pi_{\pm}(\hat n)$ are projectors onto eigenstates of $\sigma_{\hat n}=\hat n\!\cdot\!\vec\sigma$ and $g\in[0,1]$ denotes the measurement strength.

As in Sec.~\ref{gen-microscopic-model}, each system qubit $X\in\{A,B\}$ is coupled to an individual detector qubit $D_X$, initialized in $|0\rangle_{D_X}$ and measured in the computational basis $\{|0\rangle,|1\rangle\}$.  For a measurement along axis $\hat n$, we choose the conditional system–detector interaction
\begin{equation}
H^{X}_{SD}(\hat n)
=
J\,\Pi^{X}_{+}(\hat n)\otimes\sigma^{D_X}_{y}
+
\Bigl(\frac{\pi}{2\tau}-J\Bigr)
\Pi^{X}_{-}(\hat n)\otimes\sigma^{D_X}_{y},
\label{eq:SM_axis_Hamiltonian}
\end{equation}
with fixed interaction time $\tau$.

Because $\Pi_{+}(\hat n)\Pi_{-}(\hat n)=0$ and $\Pi_{+}(\hat n)+\Pi_{-}(\hat n)=\mathbb{I}$, the corresponding unitary evolution factorizes as
\begin{eqnarray}
U^{X}_{SD}(\hat n;\tau)
&=& e^{-i \tau H^{X}_{SD}(\hat n)}\nonumber \\
&=& \Pi^{X}_{+}(\hat n)\otimes e^{-iJ\tau\sigma_y}
+
\Pi^{X}_{-}(\hat n)\otimes e^{-i(\frac{\pi}{2}-J\tau)\sigma_y}.\nonumber
\label{eq:SM_axis_unitary}
\end{eqnarray}
Projecting the detector onto $|0\rangle$ and $|1\rangle$ gives the two Kraus operators
\begin{eqnarray}
K^{X}_{+}(\hat n)
&=&
\langle 0|U^{X}_{SD}(\hat n;\tau)|0\rangle \nonumber\\
&=&
\cos(J\tau)\,\Pi^{X}_{+}(\hat n)
+
\sin(J\tau)\,\Pi^{X}_{-}(\hat n),\nonumber
\end{eqnarray}
and
\begin{eqnarray}
K^{X}_{-}(\hat n)
&=&
\langle 1|U^{X}_{SD}(\hat n;\tau)|0\rangle\nonumber \\
&=&
\sin(J\tau)\,\Pi^{X}_{+}(\hat n)
+
\cos(J\tau)\,\Pi^{X}_{-}(\hat n).\nonumber
\end{eqnarray}
Identifying
\begin{equation}
\cos(J\tau)=\sqrt{\frac{1+g}{2}},
\qquad
\sin(J\tau)=\sqrt{\frac{1-g}{2}},
\label{eq:SM_strength_relation}
\end{equation}
one obtains the Kraus operators in the form given in Eq.~\eqref{eq:SM_Luders_form} with strength
\begin{equation}
g=\cos(2J\tau).
\label{eq:SM_g_relation}
\end{equation}
Thus, the microscopic interaction~\eqref{eq:SM_axis_Hamiltonian} reproduces precisely the measurement Kraus operators used in the main text.

We now implement the branch-dependent measurement axes defining our example.  Introducing a tilt angle $\theta$, we define
\begin{equation}
\hat n_0 = \cos\theta\,\hat x - \sin\theta\,\hat z,
\qquad
\hat n_1 = \cos\theta\,\hat x + \sin\theta\,\hat z.
\label{eq:SM_axes}
\end{equation}
Branch $0$ performs a $\sigma_x$ measurement on qubit $A$ and a $\sigma_{\hat n_0}$ measurement on qubit $B$, while branch $1$ performs $\sigma_{\hat n_1}$ on $A$ and $\sigma_x$ on $B$.  Accordingly, the system–detector Hamiltonians entering Eq.~(\ref{eq:Htot_SM}) of Sec.~\ref{gen-microscopic-model} are defined as
\begin{eqnarray}
H^{A,(0)}_{SD} &=& H^{A}_{SD}(\hat x),
\qquad
H^{B,(0)}_{SD} = H^{B}_{SD}(\hat n_0),\nonumber\\
H^{A,(1)}_{SD} &=& H^{A}_{SD}(\hat n_1),
\qquad
H^{B,(1)}_{SD} = H^{B}_{SD}(\hat x).\nonumber
\end{eqnarray}
Substituting these into the controlled total Hamiltonian from Eq.~(\ref{eq:Htot_SM}), we get
\begin{eqnarray}
H_{\rm tot}
&=&|0\rangle\!\langle 0|_{C}\otimes\Bigl(H^{A}_{SD}(\hat x)\otimes \mathbb{I}_{D_B}+\mathbb{I}_{D_A}\otimes H^{B}_{SD}(\hat n_0)\Bigr)\nonumber \\
&+&|1\rangle\!\langle 1|_{C}\otimes\Bigl(H^{A}_{SD}(\hat n_1)\otimes \mathbb{I}_{D_B}+\mathbb{I}_{D_A}\otimes H^{B}_{SD}(\hat x)\Bigr).\nonumber
\label{eq:SM_total_H_example}
\end{eqnarray}
After detector readout and postselection on the control qubit, one obtains exactly the branch-dependent Kraus operators used in the main-text protocol.

\section{Absence of maximal entanglement under readout averaging}
\label{app:avg_case_noncolinearity}
We show that the readout-averaged state $\rho_{AB}$ cannot be maximally entangled. Since maximal entanglement requires purity, the argument proceeds by showing that the purity condition itself forces $\rho_{AB}$ to be a product state---a contradiction. Purity of $\rho_{AB}$ requires all states $|\psi_{ij}\rangle$ to be mutually proportional. To analyze when this can hold, fix a readout $i$ on subsystem $A$ and consider different readouts $j$ on subsystem $B$. The states $|\psi_{ij}\rangle$ share the same branch-induced vectors on $A$, namely $K_i^A|\phi_A\rangle$ and $M_i^A|\phi_A\rangle$, while the contribution on $B$ depends on $j$. If these two vectors are linearly independent, varying $j$ changes the direction of $|\psi_{ij}\rangle$, violating proportionality; hence proportionality enforces colinearity of the readout-conditioned vectors on subsystem $B$ within each branch. We show below that, for the Kraus operators considered, such colinearity occurs only when $|\phi_B\rangle$ is an eigenstate of the measured observable, in which case both branches act on $|\phi_B\rangle$ by scalar multiplication and every $|\psi_{ij}\rangle$ factorizes across $B$. If, on the other hand, $K_i^A|\phi_A\rangle$ and $M_i^A|\phi_A\rangle$ are already colinear, the state factorizes across $A$ directly. In both cases, the pure state---were it to exist---would be a product state, contradicting maximal entanglement.

Consider a single subsystem $X\in\{A,B\}$ and a generalized measurement along an axis $\hat n$, described by the Kraus operators (\ref{eq:SM_Luders_form}). Writing the input state in the eigenbasis of $\hat n\cdot\vec\sigma$ as
\begin{equation}
|\phi_X\rangle=c_+|\hat n,+\rangle+c_-|\hat n,-\rangle,
\end{equation}
one finds
\begin{align}
L_+(g,\hat n)|\phi_X\rangle
&=a c_+|\hat n,+\rangle+b c_-|\hat n,-\rangle,\\
L_-(g,\hat n)|\phi_X\rangle
&=b c_+|\hat n,+\rangle+a c_-|\hat n,-\rangle.
\end{align}
If these two vectors are proportional,
\begin{equation}
L_+(g,\hat n)|\phi_X\rangle=\eta L_-(g,\hat n)|\phi_X\rangle,
\end{equation}
then comparison of coefficients gives
\begin{equation}
a c_+ = \eta b c_+,\qquad
b c_- = \eta a c_-.
\end{equation}
For $g\neq 0$, one has $a\neq b$, and the above equations cannot be satisfied simultaneously unless $c_+c_-=0$. Thus, the two readout-conditioned vectors are proportional only when $|\phi_X\rangle$ is an eigenstate of the measured observable $\hat n\cdot\vec\sigma$.

We now apply this result to the readout-averaged state. For the averaged state to be pure, all states $|\psi_{ij}\rangle$ must be mutually proportional. Fixing a readout $i$ on subsystem $A$, we compare the states corresponding to different readouts on subsystem $B$,
\begin{align}
|\psi_{i+}\rangle &=
\alpha K_i^A|\phi_A\rangle K_+^B|\phi_B\rangle
+\beta M_i^A|\phi_A\rangle M_+^B|\phi_B\rangle,\\
|\psi_{i-}\rangle &=
\alpha K_i^A|\phi_A\rangle K_-^B|\phi_B\rangle
+\beta M_i^A|\phi_A\rangle M_-^B|\phi_B\rangle.
\end{align}

If the two branch-induced vectors on subsystem $A$, namely
$K_i^A|\phi_A\rangle$ and $M_i^A|\phi_A\rangle$, are linearly independent, then proportionality of $|\psi_{i+}\rangle$ and $|\psi_{i-}\rangle$ requires that the corresponding vectors on subsystem $B$ be proportional within each branch with the same proportionality constant. In particular,
\begin{equation}
K_+^B|\phi_B\rangle = \Lambda_i K_-^B|\phi_B\rangle,
\qquad
M_+^B|\phi_B\rangle = \Lambda_i M_-^B|\phi_B\rangle.
\end{equation}
Thus, the two readout-conditioned vectors of each measurement on subsystem $B$ are colinear. From the local result above, this is possible, for $g\neq 0$, only if $|\phi_B\rangle$ is an eigenstate of the measured observable in each branch. In that case, the Kraus operators act on $|\phi_B\rangle$ only by scalar multiplication, and the resulting state factorizes across subsystem $B$. 
If, on the other hand, $K_i^A|\phi_A\rangle$ and $M_i^A|\phi_A\rangle$ are colinear, then the state already factorizes across subsystem $A$ for that readout sector. Since all $|\psi_{ij}\rangle$ must be proportional, the common state is again a product state.

The same reasoning applies with $A$ and $B$ interchanged. Therefore, purity of the readout-averaged state requires factorization on at least one subsystem: any pure readout-averaged state is necessarily a product state, and hence carries no entanglement. Maximal entanglement after averaging over detector readouts is therefore impossible.


\section{Concurrence of the readout-averaged state}
\label{app:concurrence}

Here, we derive Eq.~(\ref{concurrence-avg-regime}) for the concurrence of the readout-averaged state. Define
\begin{equation}
\rho_{AB}=\frac{1}{\mathcal{N}}\sum_{i,j}|\psi_{ij}\rangle\langle\psi_{ij}|,
\end{equation}
where
\begin{equation}
\mathcal{N}=\mathrm{Tr}\left[\sum_{i,j}|\psi_{ij}\rangle\langle\psi_{ij}|\right],
\end{equation}
is the normalization factor required to have $\mathrm{Tr[\rho_{AB}]=1}$. For the two-qubit example considered in the main text, it is convenient to work in the Bell basis
\begin{equation}
\{|\Phi^+\rangle,|\Phi^-\rangle,|\Psi^+\rangle,|\Psi^-\rangle\},
\end{equation}
where
\begin{equation}
|\Phi^\pm\rangle=\frac{|00\rangle\pm |11\rangle}{\sqrt{2}},
\qquad
|\Psi^\pm\rangle=\frac{|01\rangle\pm |10\rangle}{\sqrt{2}}.
\end{equation}
In this basis, one obtains
\begin{equation}
\rho_{AB}
=
\frac{1}{16\mathcal{N}}
\begin{pmatrix}
x & 0 & x & y \\
0 & z & 0 & 0 \\
x & 0 & x & y \\
y & 0 & y & w
\end{pmatrix},
\end{equation}
with
\begin{align}
x &= g^2(1-\cos\theta)^2,\\
y &= -g^2(1-\cos\theta)\sin\theta,\\
z &= \left[4\left(1-\sqrt{1-g^2}\right)-g^2\right]\sin^2\theta,\\
w &= g^2 \sin^2\theta,
\end{align}
and
\begin{equation}
\mathcal{N}
=
\frac{1}{8}
\left[
g^2(1-\cos\theta)^2
+2(1-\sqrt{1-g^2})\sin^2\theta
\right].
\end{equation}
The spin-flipped state entering Wootters' construction is
\begin{equation}
\tilde{\rho}_{AB}
=
(\sigma_y\otimes\sigma_y)\rho_{AB}^*
(\sigma_y\otimes\sigma_y).
\end{equation}
Let $\lambda_i$ denote the square roots of the eigenvalues of
$\rho_{AB}\tilde{\rho}_{AB}$, ordered as
$\lambda_1\geq\lambda_2\geq\lambda_3\geq\lambda_4$. A direct evaluation gives
\begin{align}
\lambda_1
&=
\frac{\left[4(1-\sqrt{1-g^2})-g^2\right]\sin^2\theta}{16\mathcal{N}},\\
\lambda_2
&=
\frac{g^2\sin^2\theta}{16\mathcal{N}},
\end{align}
with $\lambda_3=\lambda_4=0$. Therefore,
\begin{align}
\mathcal{C}_{\rm avg}(g,\theta)
&=
\max\{0,\lambda_1-\lambda_2-\lambda_3-\lambda_4\} \nonumber\\
&=
\frac{g^2(1+\cos\theta)}
{4(1+\sqrt{1-g^2})-g^2(1-\cos\theta)}.
\end{align}
This is Eq.~(\ref{concurrence-avg-regime}) of the main text.

\section{Bell nonlocality and boundary of the nonlocal region}
\label{app:bell-boundary}

We determine the region of Bell nonlocality of the readout-averaged state $\rho_{AB}$ using the Horodecki criterion~\cite{HORODECKI1995340}. For any two-qubit state, violation of the CHSH inequality occurs if and only if
\begin{equation}
\gamma(g,\theta)=\mu_1+\mu_2 > 1,
\end{equation}
where $\mu_1$ and $\mu_2$ are the two largest eigenvalues of the matrix $U=V^T V$. The matrix $V$ is defined by
\begin{equation}
V_{mn} = \mathrm{Tr}\left[\rho_{AB}\, (\sigma_m \otimes \sigma_n)\right],
\end{equation}
with $\sigma_i$ the Pauli operators.

For the state $\rho_{AB}$ obtained in the main text, one finds that $V$ takes the form
\begin{equation}
V =
\begin{pmatrix}
t & 0 & r \\
0 & u & 0 \\
-r & 0 & u
\end{pmatrix},
\end{equation}
where
\begin{align}
t &= \frac{g^2 (1 - \cos\theta)^2 - 2 (1 - \sqrt{1-g^2}) \sin^2\theta}
{g^2 (1 - \cos\theta)^2 + 2 (1 - \sqrt{1-g^2}) \sin^2\theta},\\
r &= \frac{g^2 (1 - \cos\theta)\sin\theta}
{g^2 (1 - \cos\theta)^2 + 2 (1 - \sqrt{1-g^2}) \sin^2\theta},\\
u &= \frac{(1 - \sqrt{1-g^2})^2 \sin^2\theta}
{g^2 (1 - \cos\theta)^2 + 2 (1 - \sqrt{1-g^2}) \sin^2\theta}.
\end{align}

The matrix $U$ then takes the form
\begin{equation}
U=
\begin{pmatrix}
t^2 + r^2 & 0 & r(t-u) \\
0 & u^2 & 0 \\
r(t-u) & 0 & r^2 + u^2
\end{pmatrix}.
\end{equation}
Its eigenvalues are
\begin{align}
\nu_1 &= \frac{4(1+\cos\theta)^2}
{\left(3+\sqrt{1-g^2}+\cos\theta-\sqrt{1-g^2}\cos\theta\right)^2},\\
\nu_2 &= \frac{(1-\sqrt{1-g^2})^2 (1+\cos\theta)^2}
{\left(3+\sqrt{1-g^2}+\cos\theta-\sqrt{1-g^2}\cos\theta\right)^2},\\
\nu_3 &= \frac{4(\sqrt{1-g^2}-\cos\theta)^2}
{\left(3+\sqrt{1-g^2}+\cos\theta-\sqrt{1-g^2}\cos\theta\right)^2}.
\end{align}
It follows immediately that $\nu_1 \geq \nu_2$ for all $g$ and $\theta$. The ordering between $\nu_2$ and $\nu_3$ depends on the parameters $g$ and $\theta$. The crossover between these two eigenvalues is obtained from $\nu_2=\nu_3$, which gives
\begin{equation}
\cos\theta = \frac{3\sqrt{1-g^2}-1}{3-\sqrt{1-g^2}}.
\end{equation}
The Horodecki quantity is therefore
\begin{equation}
\gamma(g,\theta) = \mu_1 + \mu_2 = \nu_1 + \max(\nu_2,\nu_3).
\end{equation}

In the region where $\nu_3\geq\nu_2$, direct substitution shows that $\nu_1+\nu_3\leq 1$; hence this branch does not lead to Bell violation. Therefore, the boundary of the Bell-nonlocal region is determined by the condition
\begin{equation}
\mu_1 + \mu_2 = 1,
\end{equation}
where $\mu_1=\nu_1$ and $\mu_2=\nu_2$. Substituting the expressions for $\nu_1$ and $\nu_2$, this condition reduces to
\begin{equation}
\frac{(1+\cos\theta)^2\left[4+(1-\sqrt{1-g^2})^2\right]}
{\left(3+\sqrt{1-g^2}+\cos\theta-\sqrt{1-g^2}\cos\theta\right)^2}
= 1.
\end{equation}

Solving this equation for $\theta$ yields the boundary curve
\begin{equation}
\theta_B(g)
=
\arccos\left[
\frac{
3+\sqrt{1-g^2}
-\sqrt{4+\left(1-\sqrt{1-g^2}\right)^2}
}{
\sqrt{4+\left(1-\sqrt{1-g^2}\right)^2}
-1+\sqrt{1-g^2}
}
\right].
\end{equation}
The Bell-nonlocal region is thus given by $0<\theta<\theta_B(g)$ (with $\theta=0$ understood as a limiting point).

\end{document}